# A Radio Frequency Channelizer based on Cascaded Integrated Micro-ring Resonator Optical Comb Sources and Filters

Xingyuan Xu[1], Jiayang Wu[1], Thach G. Nguyen[2], Sai T. Chu[3], Brent E. Little[4], Roberto Morandotti[5,6,7], Arnan Mitchell[2], and David J. Moss[1]

[1]Centre for Micro-Photonics, Swinburne University of Technology, Hawthorn, VIC 3122, Australia.
[2]School of Engineering, RMIT University, Melbourne, VIC 3001, Australia.
[3]Department of Physics and Material Science, City University of Hong Kong, Tat Chee Avenue, Hong Kong, China.
[4]State Key Laboratory of Transient Optics and Photonics, Xi'an Institute of Optics and Precision Mechanics, Chinese Academy of Science, Xi'an, China.
[5]INSR-Énergie, Matériaux et Télécommunications, 1650 Boulevard Lionel-Boulet, Varennes, Québec, J3X 1S2, Canada
[6]National Research University of Information Technologies, Mechanics and Optics, St. Petersburg, Russia
[7]Institute of Fundamental and Frontier Sciences, University of Electronic Science and Technology of China, Chengdu 610054, China.

*Abstract*
We report a broadband RF channelizer based on an integrated optical frequency Kerr micro-comb source, with an RF channelizing bandwidth of 90 GHz, a high RF spectral slice resolution of 1.04 GHz, and experimentally verify the RF performance up to 19 GHz. This approach to realizing RF channelizers offers reduced complexity, size, and potential cost for a wide range of applications to microwave signal detection.

## I. INTRODUCTION

Detecting and analyzing microwave signals with large instantaneous bandwidth is essential for a wide range of applications in satellite communications, electronic warfare, and radar systems [1-10]. RF channelizers, which can slice the RF spectrum into frequency channels with digital-compatible bandwidths [11], serve as one of the most attractive approaches for RF signal detection and analysis. Conventional RF channelizers typically employ a bank of RF filters, thus are subjected to limitations of the electrical bandwidth bottleneck. Photonic techniques are promising solutions for RF channelizers, offering large bandwidths and strong immunity to electromagnetic interference.

Early approaches to photonic RF channelizers, where RF signals were transmitted over single optical wavelengths and then physically split for spectral channelizing, were achieved via diffraction gratings [12], acousto-optic crystals [13], fiber gratings [14], and integrated devices [15]. However, those approaches required a large set of narrow, spectrally dense and precisely centered filter banks, hence suffering from limitations in spectral slice resolution, available number of channels and footprint.

Recent approaches have been demonstrated based on multicasting RF signals onto multiple optical wavelengths via stimulated Brillouin scattering [16-18], parametric processes in nonlinear fiber [19, 20], spectrally sliced incoherent sources [21], discrete laser arrays [22], or electro-optic modulator-based frequency combs [11, 23, 24]. Other approaches relying on wavelength scanning structures [25] or dispersive Fourier transformations [26] have also been proposed, but all of these face limitations including channel number and spectral slice resolution, as well as increased complexity and potential cost because of the need for many discrete components, such as external RF sources and mode-locked lasers. Integrated micro-comb sources [27-31], particularly those based on CMOS-compatible platforms [32-40], offer many advantages for broadband RF channelizers compared with conventional multi-wavelength sources, such as a much higher number of wavelengths [41-50] and greatly reduced footprint and complexity.

Here, we report a broadband RF channelizer based on an integrated optical frequency comb source. By using an on-chip Kerr comb source consisting of an active nonlinear micro-ring resonator (MRR) that yields a broadband 200 GHz-spaced comb, in combination with a passive on-chip MRR with a free spectral range (FSR) of 49 GHz and Q factor of $1.55 \times 10^6$, we realize an RF channelizer with a large number of channels (20 over the C-band and > 60 over the C/L-bands) and an RF bandwidth approaching 100 GHz. We verify the RF performance experimentally at frequencies up to 19 GHz, achieving a high spectral slice resolution of 1.04 GHz. In addition to high RF performance, this approach offers a reduced footprint, lower complexity, and potentially lower cost.

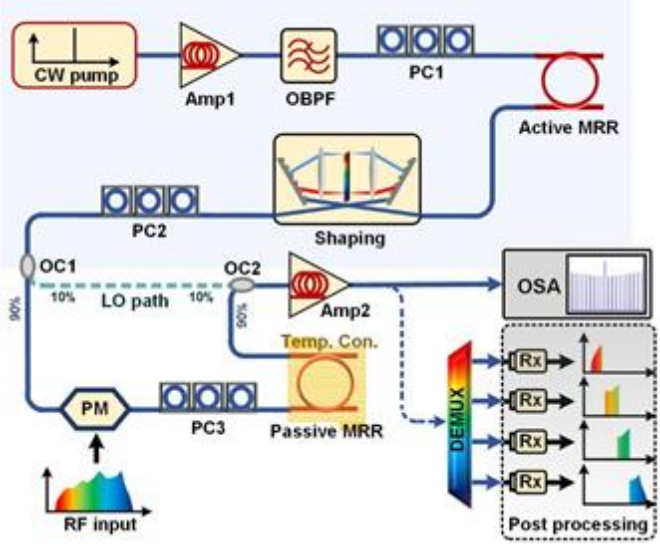

Fig. 1. Schematic diagram of the broadband RF channelizer based on an integrated optical comb source. Amp: erbium-doped fibre amplifier. OBPF: optical bandpass filter. PC: polarization controller. MRR: micro-ring resonator. OC: optical coupler. PM: phase modulator. Temp. Con.: temperature controller. DEMUX: de-multiplexer. Rx: Receiver. OSA: Optical spectrum analyzer.

## II. PRINCIPLE

Figure 1 shows a diagram of the broadband RF channelizer. Kerr combs were generated in an integrated active MRR pumped by a tunable CW laser and amplified by an erbium-doped fibre amplifier. A tunable optical bandpass filter was used to suppress the amplified spontaneous emission noise from the amplifier, while a fibre polarization controller was used to optimize the power in the waveguide. When the wavelength of the pump light was tuned to one of the active MRR resonances with the pump power high enough to provide sufficient parametric gain, optical parametric oscillation occurred, ultimately generating Kerr optical combs with equal spacing $\delta_{OFC}$ (~ 200 GHz). In the wavelength range of the channelizer, the frequency of the $k_{th}$ ($k$=2, 3, 4, …) comb line is

$$f_{OFC}(k) = f_{OFC}(1) + (k-1)\delta_{OFC} \qquad (1)$$

where $f_{OFC}(1)$ is the frequency of the first comb line used on the red side. The Kerr combs were then shaped (flattened) by a Waveshaper and directed to a phase modulator (PM), where the input broadband RF signal was multicast onto each wavelength. Finally, the 200 GHz comb lines, each imprinted with the RF spectrum to be measured, were spectrally sampled, or sliced, by a separate passive high-Q MRR with an FSR of 49 GHz, where we used every $4_{th}$ resonance for spectral slicing, corresponding to an equivalent spacing $\delta_{MRR}$ of ~196 GHz, and so the RF spectrum on the different 200 GHz-spacing comb lines are sampled sequentially with approximately a 4 GHz shift between adjacent comb lines. The output channelized RF frequencies can be given by:

$$
\begin{aligned}
f_{RF}(k) &= f_{MRR}(k) - f_{OFC}(k) \\
&= [f_{MRR}(1) - f_{OFC}(1)] + (k-1)(\delta_{MRR} - \delta_{OFC})
\end{aligned}
\qquad (2)
$$

where $f_{RF}(k)$ is the $k_{th}$ channelized RF frequency, $f_{MRR}(k)$ is the $k_{th}$ centre frequency of the filtering MRR. Here, $[f_{MRR}(1) - f_{OFC}(1)]$ is the relative spacing between the first comb line and adjacent filtering resonance, corresponding to the offset of the channelized RF frequencies, and ($\delta_{OFC} - \delta_{MRR}$) corresponds to the channelized RF frequency step between adjacent wavelength channels. In principle, the RF channels can readily be separated by an optical demultiplexer such as an arrayed waveguide grating with a channel spacing matching the FSR (or a multiple) of the 49

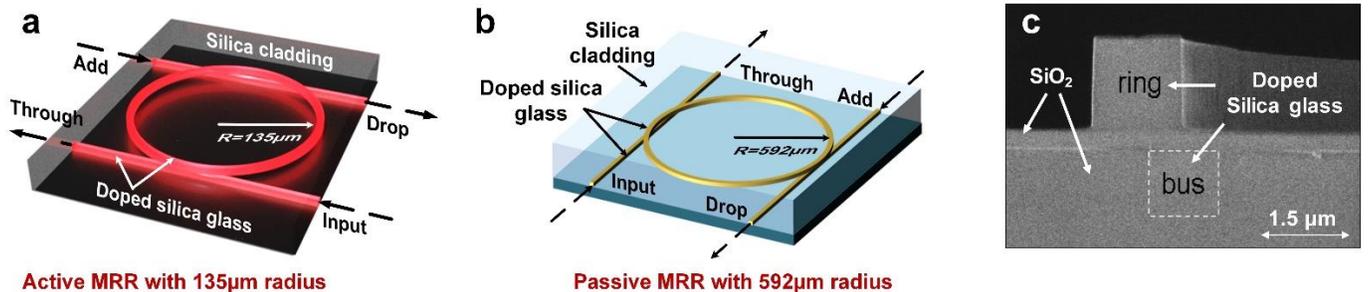

Fig. 2. Schematic illustration of the (a) 200 GHz-FSR MRR and (b) 49 GHz-FSR MRR. (c) SEM image of the cross-section of the 200 GHz MRR before depositing the silica upper cladding.

GHz MRR. Then RF signals can be detected via homodyne detection, where the flattened comb lines are first separated out before phase modulation and serve as the local oscillators (LO), then coupled together with the channelized RF optical sidebands and coherently detected for post processing. In this work, however, we measured the broadband optical spectrum to verify the feasibility of our approach, deferring channel demultiplexing and RF detection to a future work. We note that owing to the large number of wavelengths provided by the on-chip micro-comb, the RF channelizer featured a large number of wavelength channels, thus enabling broadband RF channelizing with digital-compatible channel bandwidths.

## III. EXPERIMENT

The MRRs (Fig. 2(a, b)) were fabricated in a high-index doped silica glass platform using CMOS compatible fabrication processes. First, high-index (n = ~1.70 at 1550 nm) doped silica glass films were deposited using standard plasma enhanced chemical vapour deposition (PECVD), then patterned photo-lithographically and etched via reactive ion etching to form waveguides with exceptionally low surface roughness. Finally, silica glass (n = ~1.44 at 1550 nm) was deposited via PECVD as an upper cladding. The advantages of our platform for nonlinear OPOs include ultra-low linear loss (~0.06 dB $\cdot$ cm$^{-1}$), a moderate nonlinearity parameter (~233 W$^{-1}\cdot$ km$^{-1}$), and in particular a negligible nonlinear loss up to extremely high intensities (~25 GW $\cdot$ cm$^{-2}$) [28]. The low linear loss enabled the ring resonators to achieve $Q$ factors > 1×10$^6$. A scanning electron microscope (SEM) image of the cross-section of the 200 GHz MRR before depositing the SiO$_2$ upper cladding is shown in Fig. 2(c). The radii of the active MRR (for comb generation) and the passive MRR (for spectral slicing) were ~135 μm and ~ 592 μm, corresponding to FSRs of ~1.6 nm (~200 GHz) and ~0.4 nm (~49 GHz), respectively. After packaging the device with fibre pigtails, the through-port insertion loss was ~3.5 dB for the 200 GHz MRR and ~1.5 dB for the 49 GHz MRR.

To generate Kerr combs, the pump power was boosted to ~500 mW and the wavelength was swept from blue to red. When the pump wavelength was tuned close to the resonance of the active MRR, primary combs were generated (Fig. 3(a)). Signal and idler lines were generated in the S-band and L-bands respectively, and the spacing between the signal and pump was 19 FSR, or 3.8 THz, determined by the parametric gain curve. When the detuning between the pump wavelength and the adjacent resonance was further changed, the parametric gain lobes broadened and secondary comb lines with a spacing equal to the FSR appeared via cascaded four wave mixing. Finally, flat Kerr combs were generated with the pump wavelength set to 1548.58 nm. As shown in Fig. 3(b–c), the resulting Kerr comb was over 200-nm wide, covering four bands (S, C, L, U) and was relatively flat over the full C and L bands, thus enabling a record large number of wavelengths (20 over the C-band and > 60 over the C/L-bands). The large Kerr comb spacing also yielded an increased Nyquist zone, corresponding to an RF bandwidth of >100 GHz. This bandwidth is extremely challenging for mode-locked lasers and externally-modulated comb sources to achieve [41–44].

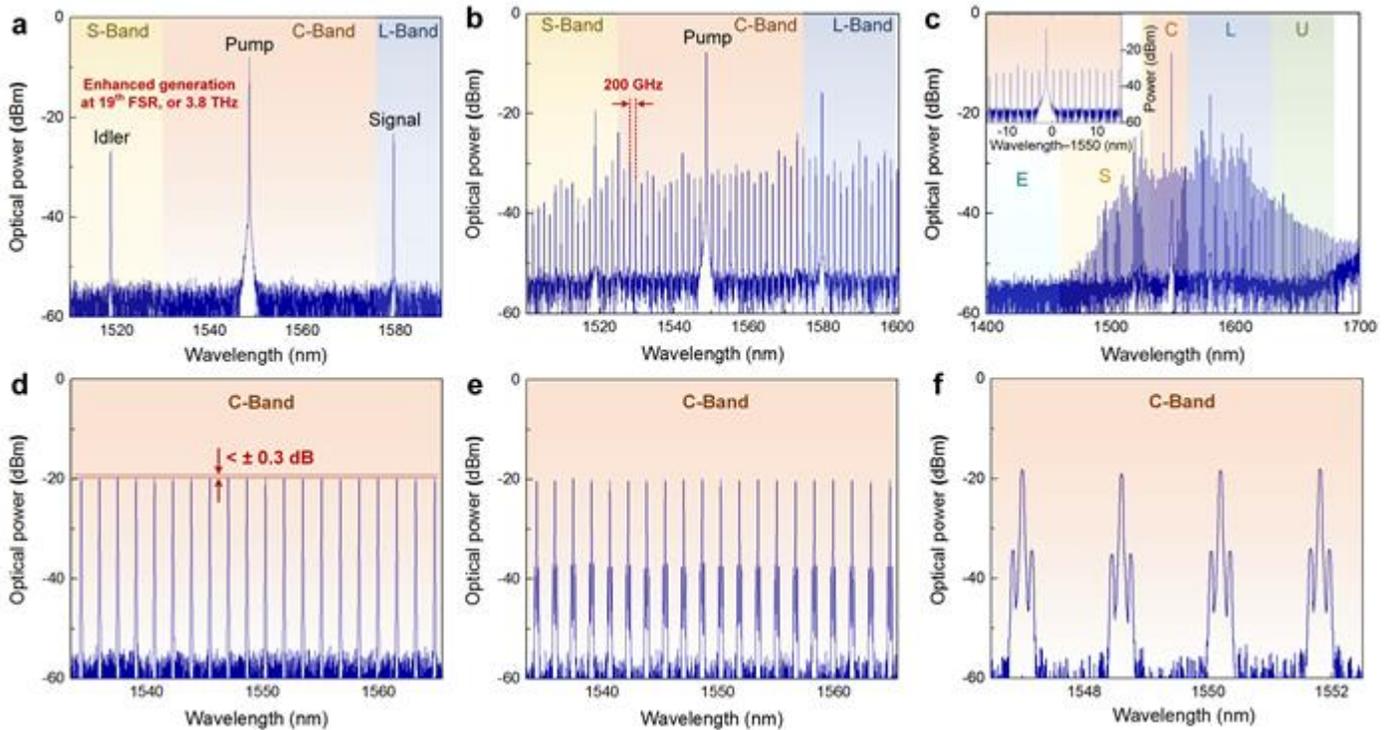

Fig. 3. Optical spectrum of (a) the primary comb, (b) the secondary comb, (c) the Kerr comb with 300 nm span, (d) the shaped optical comb for the channelizer with less than 0.5 dB unflatness, (e) 20 and (f) selected 4 comb lines modulated by RF signals.

While the spectral profile of our comb was not indicative of operation in the single cavity soliton regime, we do not foresee any potential limitations related to that, given the range of Q factors and nonlinear figure of merit that we can achieve. In addition, from theoretical analysis of what we have observed, it is evident that the Kerr comb was working under "partially coherent" conditions that features stable phase and amplitude. In fact, the spatial patterns of what we observed appear to be similar to the so called "soliton molecules" [50, 51], which feature reasonably low intensity noise comparable with that of cavity solitons.

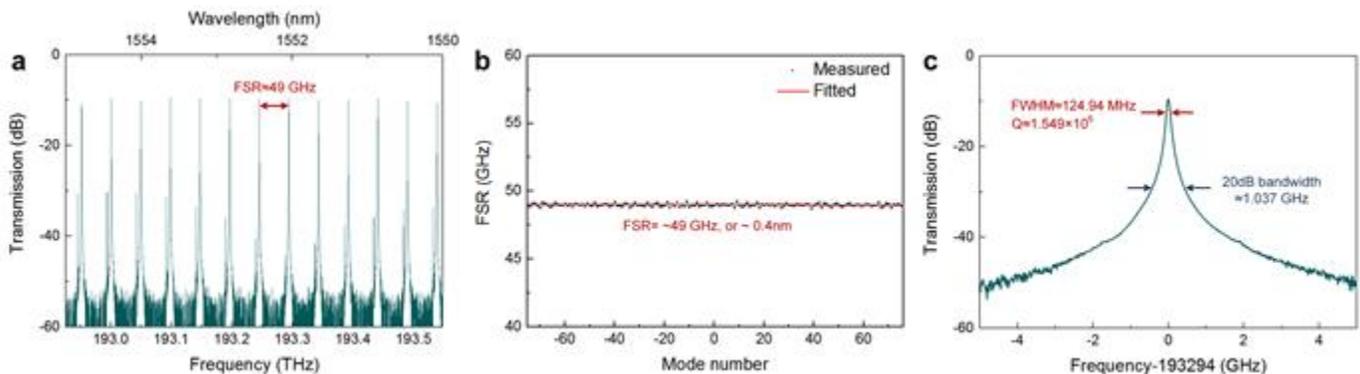

Fig. 4. Drop-port transmission spectrum of the on-chip 49 GHz MRR (a) with a span of 5 nm, (b) showing an FSR of 49 GHz, and (c) a resonance at 193.294 THz with full width at half maximum (FWHM) of 124.94 MHz, corresponding to a Q factor of $1.549 \times 10^6$.

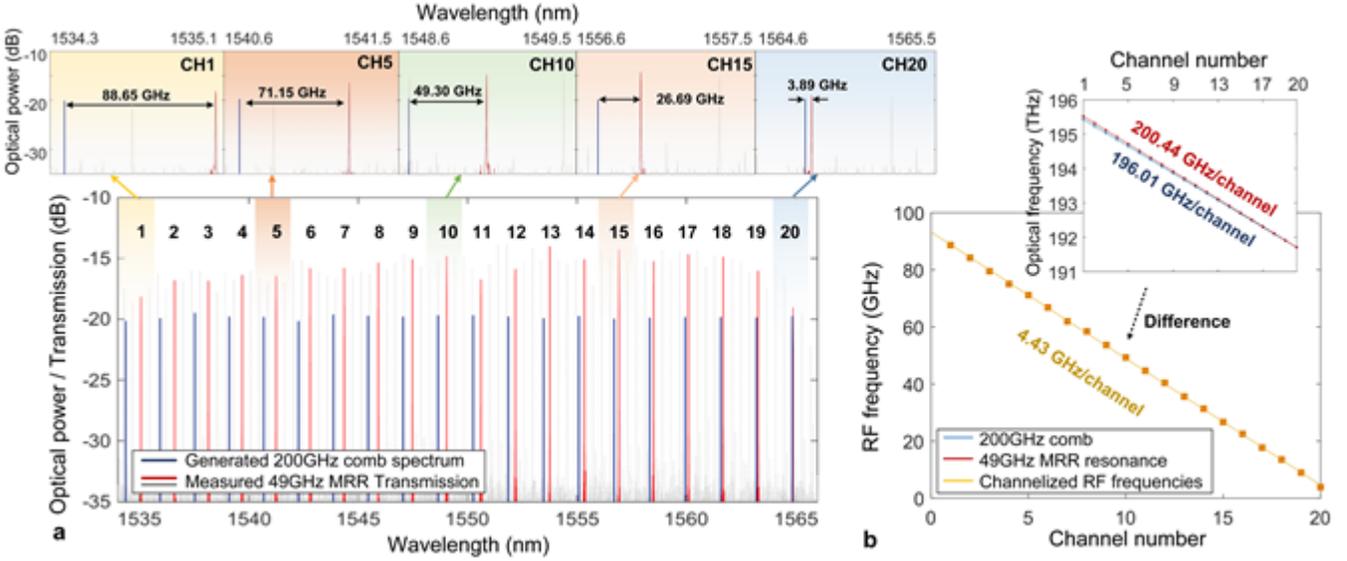

Fig. 5. (a) The measured optical spectrum of the 200GHz micro-comb and transmission of 49GHz MRR. Zoom-in views of the channels with different channelized RF frequencies. (b) Extracted channelized RF frequencies, inset shows corresponding optical frequencies of the comb lines and spectral slicing resonances.

We note that since the comb source serves as both the optical carrier for RF sideband generation as well as the LO for RF receiving, the channelized RF optical sidebands and LOs are inherently coherent, and thus strictly phase-locked comb lines are not necessary—which, in fact, is an advantage when comparing with RF coherent receivers [52]. Neither is the coherence between adjacent WDM channels required since they are not beating with each other.

The 20 channels of the Kerr comb over the C-band were then shaped by a Waveshaper (Finisar 4000S) to get uniform channel weights. A feedback control path was introduced to read and shape the comb lines' power accurately, which were first detected by an optical spectrum analyser and then compared with the ideal weights (uniform in our case). This allowed us to generate an error signal to feed back into the waveshaper in order to calibrate the system and achieve accurate comb shaping and flatness of the comb envelope to within ± 0.3 dB (Fig. 3(d)). The phase modulator was then used to multicast the input RF signal onto each comb line. Figures. 3(e–f) show the optical spectra of the flat comb lines modulated by the RF signal. The number of equalized channels (20) was limited to the C-band by the bandwidth of the waveshaper. In principle the number of wavelengths could easily be increased up to 60 by using a combined C+L band waveshaper. Next, the multicast RF signal on each wavelength channel was spectrally sliced by the passive 49 GHz MRR. Figures. 4(a–b) show the drop-port transmission of the MRR over a 5nm range, showing an FSR of ~49 GHz (~0.4nm) over 152 modes (labelling the mode at 193.294 THz as 0). The transmission of a single resonance (Fig. 4(c)) shows a full width at half maximum (FWHM) of 124.94 MHz, corresponding to a Q factor of $1.549{\times}10^{6}$, and a 20dB bandwidth of 1.04 GHz, which determines the RF resolution. This RF resolution is compatible with state-of-the-art analog-to-digital converters [3], thus leading to digital-compatible broadband RF channelization.

To illustrate the principle of operation of our RF channelizer, we plot the measured 200 GHz comb spectrum and transmission of the 49 GHz passive MRR (highlighting every 4th resonance that closely aligns with the 200 GHz comb) in the same figure (Fig. 5(a)). Zoom-in views in Fig. 5(a) show that over the 20 comb channels that are generated across the C-band, the relative shift between the 200 GHz comb lines and every 4th resonance of the 49 GHz MRR, which corresponds to the RF spectral sliced frequencies, ranges from 3.89 to 88.65 GHz. The frequencies and relative spacing of the 200 GHz comb and 49 GHz MRR are linearly fit and shown in Fig. 5(b), where we see that the fit comb spacing is $\delta_{OFC}$ = 200.44 GHz and the fit FSR of the spectrally sliced resonances (i. e., $\delta_{MRR}$ is the frequency spacing between every 4th resonance of the 49 GHz MRR) is $\delta_{MRR}$ = 196.01 GHz, thus resulting in a channelized RF frequency step between adjacent channels ($\delta_{OFC}-\delta_{MRR}$) of 4.43 GHz.

We measured the RF performance of the channelizer up to 20 GHz – the limit of our equipment. This only required 4 wavelengths or channels of the 20 channels shown in Figure 5, corresponding to four RF frequencies ranging from 1.7 to 19.0 GHz. We measured the output of the 49 GHz passive MRR with an optical spectrum analyzer (Fig. 6), showing that optical RF sidebands with different RF frequencies had been channelized at different optical wavelengths. As Fig. 6(a) shows, optical-carrier RF tones at frequencies of 1.7 GHz, 6.3 GHz, 11.2 GHz, and 15.8 GHz were output at four different wavelength channels, corresponding to an RF frequency step between adjacent wavelength channels of ~ 4.8 GHz, in agreement with Fig. 5. The power leakage of channel 4 was mainly due to spurious optical carrier tones, which could be reduced by adopting an optimized modulation format such as carrier-suppressed single-sideband

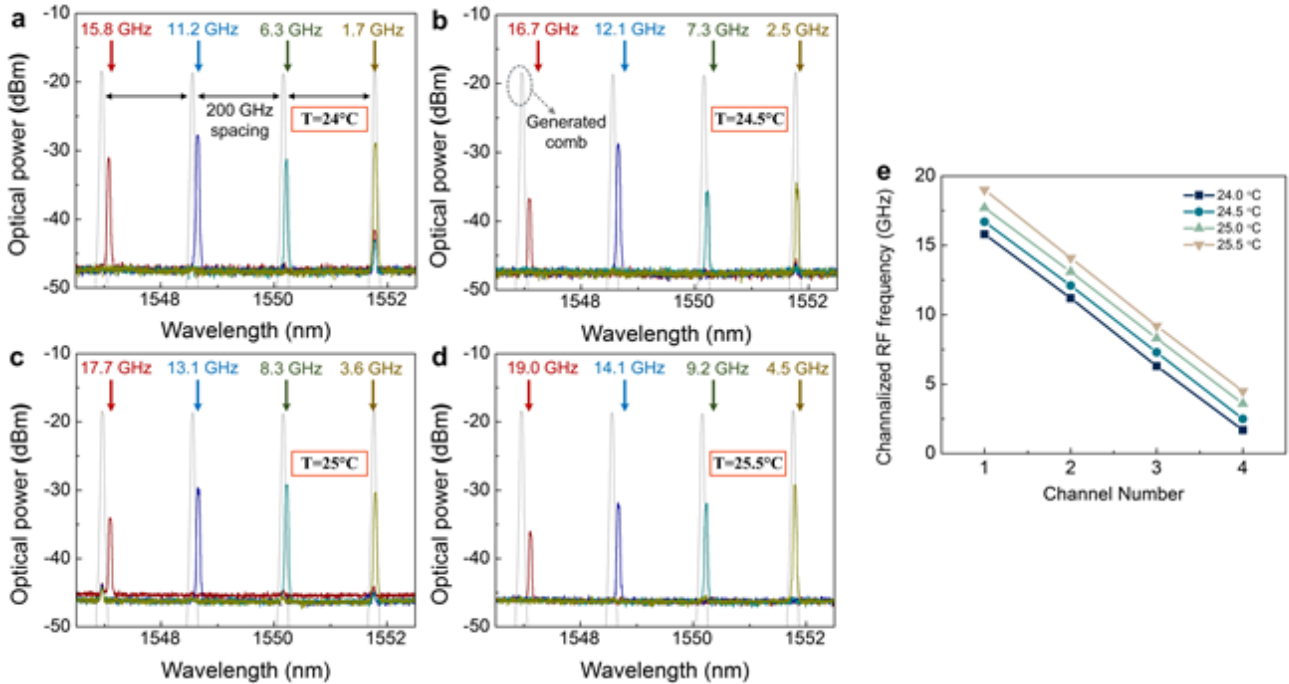

Fig. 6. RF response. Measured optical spectrum of 49 GHz MRR's output with different input RF frequencies, with the temperature of the 49 GHz MRR set to (a) 24.0℃, (b) 24.5℃, (c) 25.0℃, and (d) 25.5℃. (e) Channelized RF frequencies at different wavelength channels with different temperatures.

modulation.

In order to make full use of the RF resolution of our system (~1 GHz) and bridge the gap formed by the ~ 4.8 GHz RF step size, we used temperature tuning [31, 53] of the passive MRR to continuously control the relative spacing between the source and filtering MRRs' resonances ($f_{MRR}(1) - f_{OFC}(1)$), where a millisecond time-scale thermal response time is expected. Figures. 6(a–d) show the output optical spectra with temperatures ranging from 24.0℃ to 25.5℃. As reflected in Fig. 6(e), the offset of the channelized RF frequencies increased with a slope of ~1GHz/℃. By combining the fine resolution temperature tuning with the 4.8 GHz RF increment across 60 channels over the C/L-bands, this device is capable of achieving an ultrawide RF bandwidth of over 100 GHz. Figure 7 shows the experimental channelized RF frequencies of the different optical channels extracted from Fig. 6, covering an RF bandwidth from 1.7 to 19.0 GHz, limited by the bandwidth of our RF modulator, verifying the effectiveness of thermal tuning to generate the broad operation bandwidth of our approach. Figure 8 shows the extinction ratio (ER) of the wavelength channels (i.e., the optical power ratio of the signal channel to the maximum of other channels) extracted from Figs. 6(a–d), showing up to 20 dB ER. The non-uniformity in optical power of the different channels mainly arose from the amplifier spectral gain profile, and this can be mitigated by using gain flattening filters. The extinction ratio was also limited by excess noise of the second EDFA when adapting the optimized RF modulation format to increase the OSNR. In this case, using lower noise EDFAs would enhance the performance. Moreover, the employed thermal controllers of the active and passive MRRs featured an accuracy of 0.01°C, corresponding to an error of ~50 MHz for the channelized RF frequencies, which can be eliminated via precise thermal control or a feedback loop to achieve high performance in practical applications.

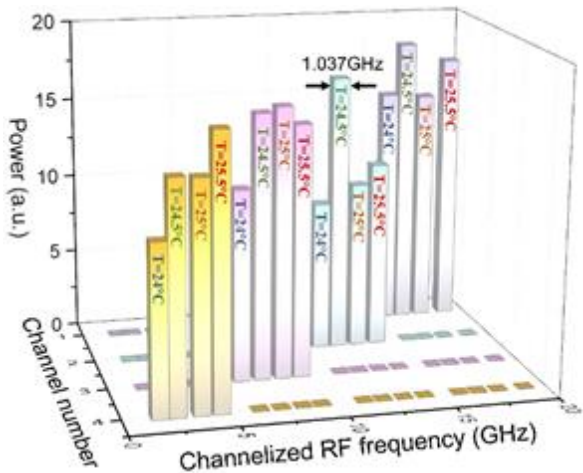

Fig. 7. Channelized RF frequencies at different channels.

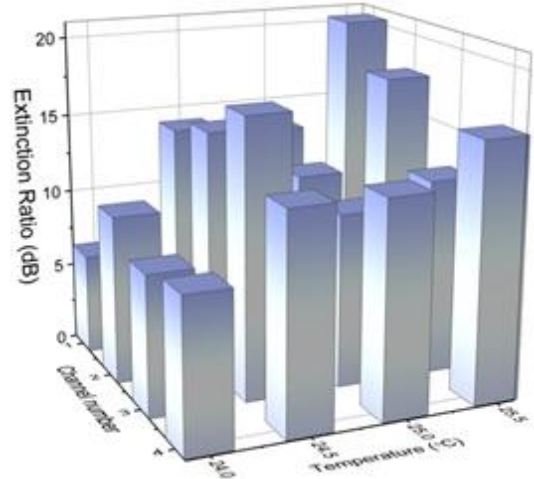

Fig. 8. Extracted extinction ratio of channelized RF signals.

Although employing thermal tuning brought challenges for simultaneous full RF spectrum channelizing, it nonetheless allowed us to demonstrate the potential and novelty of our approach. In addition, we note that thermal tuning is not necessary, provided the RF frequency step between adjacent wavelength channels (i.e., the FSR difference between the active MRR and the passive MRR) is equal to the spectral slicing resolution (i.e, the 20dB bandwidth of the passive MRR), which can be achieved via precise lithographic control of the FSRs of the MRRs. For example, given $\delta_{OFC}$=200 GHz, then designing the passive MRR such that $\delta_{MRR}$=199 GHz, still with a 1 GHz 20dB-bandwidth, would result in a channelized RF frequency step of 1 GHz and a spectral slicing resolution of 1 GHz at the same time. Thus, the full RF spectra could be simultaneously channelized with 1GHz resolution without the need for temperature tuning. While the RF operation bandwidth, given by the product of channel number and resolution, would be slightly less than the device reported here, at 60 GHz for 60 wavelengths (channels) across the C/L bands, this could be increased by using a smaller FSR comb (eg., 100 GHz vs. 200 GHz).

Finally, we note that our scheme generates RF output based on homodyne detection, whereas down-conversion of the channelized RF signal would require heterodyne detection where a second optical frequency comb with specially designed FSR and offset is required. Recent advances in dual-comb spectroscopy provide new possibilities for that [23, 54-56]. By employing integrated dual micro-combs with specially designed FSRs and offsets for heterodyne detection [57-59], broadband RF signals can be channelized and down-converted into digital bandwidths for direct analog-to-digital conversion and post processing, thus offering a highly competitive approach for integrated photonic RF receivers.

## IV. CONCLUSION

We demonstrate a broadband RF channelizer based on a CMOS-compatible integrated optical frequency comb source. Broadband 200 GHz-spacing Kerr combs with a large number of comb lines were generated by an active on-chip micro-ring resonator, providing a record large number of wavelength channels as well as an RF operation bandwidth of about 100 GHz. We demonstrate the broadband channelization of RF frequencies from 1.7 GHz to 19 GHz with a high spectral slice resolution of 1 GHz via thermal tuning of a passive on-chip MRR with a Q factor of $1.549 \times 10^6$. This micro-comb based RF channelizer is highly attractive for achieving broadband RF channelization with large channel numbers, high resolution, small footprint, and potentially low cost.